\documentclass[journal,twoside,web]{IEEEtran}
\usepackage{tmi}
\usepackage{cite}
\usepackage{dsfont}
\usepackage{amsmath,amssymb,amsfonts}
\usepackage{bm}
\usepackage{float}
\usepackage{caption}
\usepackage{subcaption}
\usepackage{setspace}
\usepackage{algorithm}

\usepackage{amsthm}
\usepackage{algpseudocode}
\usepackage{bigints}
\usepackage{relsize}
\usepackage{braket}
\usepackage{graphicx}
\usepackage{textcomp}
\usepackage{stmaryrd}
\usepackage{array,booktabs,,dcolumn,caption}
\usepackage[colorlinks=true, allcolors=blue]{hyperref}
\theoremstyle{definition}

\newcommand{\R}{\mathbb{R}}
\newcommand{\z}{\mathbf{z}}
\newcommand{\f}{\mathbf{f}}

\newcommand\numberthis{\addtocounter{equation}{1}\tag{\theequation}}
\newcommand{\Expected}{\mathbb{E}}
\newcommand{\x}{\mathbf{x}}

\newcolumntype{d}[1]{D{.}{.}{#1}} % "decimal" column type
\def\equationautorefname~#1\null{Eq.~(#1)\null}

\makeatletter
\newcolumntype{B}[3]{>{\boldmath\DC@{#1}{#2}{#3}}c<{\DC@end}}

\DeclareMathAlphabet\mathbfcal{OMS}{cmsy}{b}{n}
\def\BibTeX{{\rm B\kern-.05em{\sc i\kern-.025em b}\kern-.08em
    T\kern-.1667em\lower.7ex\hbox{E}\kern-.125emX}}
\begin{document}
%\doublespacing

%\title{GAN it learn? Towards meaningful evaluation of generative stochastic image models in medical imaging}
%\title{GAN it learn? Assessing the ability of generative image models to learn relevant statistics}
\title{Assessing the ability of generative adversarial networks to learn canonical medical \\ image statistics}

\author{Varun A. Kelkar, Dimitrios S. Gotsis, Frank J. Brooks, Prabhat KC, Kyle J. Myers, Rongping Zeng\\ and Mark A. Anastasio
\thanks{This research was supported in part by the NIH Awards EB020604, EB023045, NS102213, EB028652, and the FDA critical path funding. Varun A. Kelkar acknowledges funding by an appointment to the Research Participation Program at the Center for Devices and Radiological Health administered by the Oak Ridge Institute for Science and Education through an inter-agency agreement between the U.S. Department of Energy and U.S. Food and Drug Administration. }
\thanks{Varun A. Kelkar and Dimitrios S. Gotsis are 
with the Department of Electrical and Computer Engineering, University of Illinois at Urbana-Champaign, Urbana, IL 61801 USA (e-mail: vak2@illinois.edu; gotsis2@illinois.edu).}
\thanks{Frank J. Brooks and Mark A. Anastasio are with 
the Department of Bioengineering, University of Illinois at Urbana-Champaign, Urbana, IL 61801 USA (e-mail: fjb@illinois.edu; maa@illinois.edu).}
\thanks{Prabhat KC and Rongping Zeng are, and Kyle J Myers (retired) was with the Center for Devices and Radiological Health, Food and Drug Administration, Silver Spring, MD 20993 USA (e-mail: Prabhat.Kc@fda.hhs.gov; drkylejmyers@gmail.com; rongping.zeng@fda.hhs.gov).}
}
\maketitle

\begin{abstract}
In recent years, generative adversarial networks (GANs) have gained tremendous popularity for potential applications in medical imaging, such as medical image synthesis, restoration, reconstruction, translation, as well as objective image quality assessment. Despite the impressive progress in generating high-resolution, perceptually realistic images, it is not clear if modern GANs reliably learn the statistics that are meaningful to a downstream medical imaging application. In this work, the ability of a state-of-the-art GAN to learn the statistics of canonical stochastic image models (SIMs) that are relevant to objective assessment of image quality is investigated. It is shown that although the employed GAN successfully learned several basic first- and second-order statistics of the specific medical SIMs under consideration and generated images with high perceptual quality, it failed to correctly learn several per-image statistics pertinent to the these SIMs, highlighting the urgent need to assess medical image GANs in terms of objective measures of image quality.
\end{abstract}


\section{Introduction}\label{sec:intro}

When developing improved medical imaging technologies, such as methods for image reconstruction, restoration, and analysis, it is crucial to objectively evaluate them via a diagnostic clinical task \cite{barrett, jha_objective_eval, sr_eval, eval_denoising}. Because a full-fledged clinical trial of rapidly developing imaging technologies often is infeasible \cite{hmi, fuli}, computer simulation studies \cite{vit} have been proposed as an alternative. In order to refine and assess any medical imaging technology via computer simulation, the nature and variability of the objects to-be-imaged must be accurately characterized. To this end, a variety of stochastic object models (SOMs) have been developed \cite{hmi, vit}; these enable simulation of random, and sufficiently realistic, digital medical objects.

{A generative model is a statistical model of an unknown data distribution that enables sampling from the data distribution via a learned representation of it.} The model is trained directly on a large sample drawn from the data distribution \cite{gdl_book}. Modern generative models learn a neural network-based mapping from a tractable distribution, such as a multivariate, independent, and identically distributed (i.i.d.) Gaussian distribution, to the intractable, high-dimensional object distribution of interest. This enables sampling from the unknown distribution, and provides the ability to perform inference. Therefore, generative models such as generative adversarial networks (GANs) are being actively investigated for applications in medical imaging, such as: image restoration \cite{gan_denoising, gancircle}, image reconstruction \cite{picgm, cs_sgm, inn_mri, iagan_sayantan}, image analysis \cite{madgan, abnormality_detection}, image-to-image translation \cite{cyclegan_medical}, data sharing \cite{gan_data_sharing} and objective image quality assessment \cite{gan_mcmc}. 

Modern generative models, such as the StyleGAN and its successors \cite{stylegan, stylegan2, stylegan3}, represent a tremendous improvement in terms of the stability, controllability, diversity, and visual quality of generated images. However, state-of-the-art GANs trained on medical image datasets have been shown to produce images that look realistic, but nevertheless contain medically impactful errors \cite{gan_data_sharing, image_translation_hallucinations, hallucinations}. Therefore, in order for GANs to be safely used in medical imaging, they must first be objectively evaluated \cite{weimin_ambientgan}, for instance, with the help of a relevant diagnostic task.

Despite tremendous improvements in the quality of the images generated by a GAN, the question of whether or not a GAN correctly approximates the statistical features important to a medical imaging application remains largely unanswered. Although mathematical summaries, such as the Wasserstein metric \cite{wgan} and negative log-likelihood \cite{gan_eval_review} are correlated with the fidelity of the trained GAN, there is no guarantee that a favorable value achieved by these measures also translates to usefulness of the trained GAN for medical imaging applications. Although perceptual measures such as the Frechet Inception distance (FID) have grown to be immensely popular, they are agnostic to the downstream task a medical image GAN may be used for \cite{fid}. Furthermore, the above mentioned measures are ensemble measures. It has been shown that individual samples drawn from the GAN may contain impactful errors despite giving satisfactory ensemble measures \cite{rucha}. Lastly, medical image distributions typically consist of multiple classes or modes, and it has been shown that may produce critical errors while producing images from a mode that is rarely seen during training \cite{gan_data_sharing}. 

The objective of this study is to assess the ability of a state-of-the-art GAN to learn the statistics of a canonical stochastic image model (SIM) that are relevant to the objective assessment of image quality (IQ), and to study how the performance assessment of GANs by task-agnostic measures such as FID score compares with the performance assessed by the medically meaningful measures identified for the canonical SIM under consideration. To this end, three canonical SIMs were identified, namely the modified clustered lumpy background model \cite{clb2}, the B-mode ultrasound speckle model \cite{insana} and the stylized two-dimensional (2D) VICTRE (S2V) model \cite{victre}. A state-of-the-art GAN architecture, namely StyleGAN2, was trained on images generated from these canonical SIMs. Statistical quantities that are meaningful and relevant to the above SIMs were computed from images from the canonical SIM as well as the images generated by the GAN. Summary measures computed from these identified statistical quantities were compared against the FID for the purpose of assessing the fidelity of the trained GAN. This work is an extension of a preliminary study conducted using an angiographic SIM \cite{gan_eval_spie}.

The remainder of this paper is organized as follows. Section \ref{sec:bkd} describes the relevant background on GANs and their evaluation, as well as the background on the SIMs used in this study. Section \ref{sec:num_methods} describes the setup for the specific numerical studies and the identification of the SIM-pertinent evaluation measures. Section \ref{sec:results} presents the results. A summary of the salient findings of this work is presented in Section \ref{sec:discussion}.
% , and a conclusion is provided in \ref{sec:conclusion}.

% Generative models, such as generative adversarial networks (GANs) are a class of models that seek to approximate unknown high-dimensional data distributions, for instance, image distributions \cite{gan_goodfellow, goodfellow_dl_book}. GANs hold promise for potential applications in medical imaging \cite{gan_mi_review}, such as unconditional medical image synthesis \cite{weimin_ambientgan, 3dstylegan}, image restoration and reconstruction \cite{inn_mri, gan_circle}, medical image translation \cite{medgan} and data augmentation \cite{gan_augmentation}. GANs have also been proposed as a tool for establishing stochastic image models (SIMs), with potential applications to objective assessment and optimization of medical imaging systems \cite{hmi, gam_mcmc}. Modern GANs such as the StyleGANs \cite{stylegan, stylegan2} are known to produce diverse, visually realistic images.

% Despite the apparent realism and diversity in images, the extent to which a GAN learns image statistics is still an ongoing topic of research \cite{gan_evaluation, gan_test}. The Frechet inception distance (FID) score, inception score (IS) and average log-likelihood are a few among a long list of metrics that have been proposed to evaluate GANs \cite{gan_evaluation}. However, these metrics are generic -- they do not address the specific issues relevant to medical imaging, and are not computed by use of those statistics that are of utility to an observer for performing a specified task. 

\section{Background}\label{sec:bkd}

\subsection{Generative adversarial networks (GANs)}
Generative adversarial networks (GANs) are a popular class of generative models that are aimed to approximate a data distribution by learning to map a sample $\z \in \R^k$ from a lower dimensional, tractable data distribution $p_{\z}$, such as the i.i.d. standard normal distribution, to a sample $\f$ from the high dimensional data distribution $p_{\f}$. 
In GANs, two networks, namely a \textit{generator network} $G : \R^k \rightarrow \R^n$ with parameters $\Theta_G$ and a \textit{discriminator network} $D : \R^n \rightarrow \R$ with parameters $\Theta_D$ are jointly trained by approximately solving the following min-max optimization problem:
\begin{align*}\label{eqn:gan_loss}
    \min_{\Theta_G} \max_{\Theta_D} ~ \Expected&_{\f \sim p_\f}  \left[ \ell(D_{\Theta_D} (\f)) \right] 
    + \Expected_{\z \sim p_\z} \left[ \ell (1 - D_{\Theta_D}(G_{\Theta_G}(\z))) \right],\numberthis{}
\end{align*}
\noindent where $\ell(\cdot)$ is a utility function used to define the objective; for instance, a popular choice being $\ell(\x) = \log(\x)$ \cite{gan_goodfellow}. The promise of a generative model such as a GAN comes from the fact that once trained, samples from the otherwise inaccessible high dimensional distribution $p_\f$ can be obtained by sampling low dimensional vectors, known as latent vectors $\z$ from $p_\z$ and computing $G(\z)$. Thus, the GAN provides a tractable representation of $p_\f$ that may find use in downstream applications in imaging science, such as image reconstruction \cite{iagan_sayantan, picgm} and image quality assessment \cite{gan_mcmc}. 

\subsection{Advanced GAN training strategies}
Under prescribed theoretical conditions, minimizing the GAN training loss described in \autoref{eqn:gan_loss} is equivalent to minimizing the empirical Jensen-Shannon (JS) divergence between the true and the estimated probability distribution functions (PDFs) of the data \cite{gan_goodfellow}. However, in practice, GAN training is known to be unstable \cite{gan_improved_training, gan_principled_training} and several strategies have been proposed to improve stability. For example, the use of different learning rates and update frequencies for the generator and discriminator weights aids in avoiding the vanishing gradients problem for the generator and premature overfitting of the discriminator \cite{gan_goodfellow, gan_review}. Novel loss functions, such as in so-called Wasserstein GANs \cite{wgan}, also help in improving the training stability. Karras, \textit{et al.} \cite{progan} proposed a strategy for scaling GANs by use of \textit{progressive training}, where both the generator and discriminator are trained on lower resolution images and are progressively grown to enable training on higher and higher resolution images. StyleGAN and its successor, StyleGAN2, introduce blocks of transformed latent vectors as inputs to different layers of the network at different resolutions, thus controlling features at different scales \cite{stylegan, stylegan2}. Although these improvements to the GAN architecture and training have cumulatively led to state-of-the-art performance in terms of diversity, controllability and realism of images generated, they are largely heuristic, and are not designed specifically to learn task-pertinent statistics of medical image distributions.

\subsection{Evaluation of generative adversarial networks}
Modern GANs, such as the StyleGAN2 \cite{stylegan2} have shown impressive performance in terms of the perceptual quality of the generated images, invertibility, and meaningful control over image semantics. However, evaluating the quality of the distribution learned by a generative model is an open problem \cite{gan_eval_review_new}. Some measures directly estimate analytical quantities and distance metrics related to the image probability density function (PDF), such as the negative log-likelihood \cite{gan_eval_review} or the Wasserstein metric \cite{wgan}. Other measures such as the perceptual path-length \cite{stylegan} analyze the nature of the manifold learned by the GAN. Motivated by subjective perceptual assessment by humans \cite{is}, perceptual evaluation measures such as the Inception score (IS) and more commonly, the Fr\'echet Inception distance (FID) score have become immensely popular \cite{is, fid}. In order to compute these scores, image features are first extracted using a pre-trained Inception network \cite{inception} and distance metrics on the extracted features are computed. 
% Given data sampled from each of the real and GAN-approximated distributions $p_r$ and $p_g$, the FID score is computed as \cite{fid}
% \begin{align*}
%     {\rm FID}(p_r, p_g) = \norm{\mu_r - \mu_g}_2^2 + {\rm Tr} \big( \Sigma_r + \Sigma_g
%     - 2(\Sigma_r \Sigma_g)^{\frac{1}{2}}  \big),\numberthis{}\label{eqn:fid}
% \end{align*}
% where $\mu_r$ and $\mu_g$ are the estimated means of the Inception features of samples drawn from $p_r$ and $p_g$ respectively, $\Sigma_r$ and $\Sigma_g$ are their respective covariance matrices, and $\rm{Tr}$ denotes the matrix trace. 
Although the FID score has shown excellent agreement with subjective visual assessments by humans \cite{fid}, it is agnostic to the downstream task a medical image GAN may be used for.
% , and it may not be suitable for evaluating the GAN with respect to the meaningful image statistics that govern task performance. 
Additionally, it is an ensemble statistic, and hence could be blind to specific errors in high-order statistics of individual images \cite{rucha}. 

The studies described below seek to assess the ability of medical image GANs to reproduce image statistics that are meaningful and pertinent to the medical stochastic image model under consideration, and to see how well traditional measures such as the FID correlate with these pertinent statistics. In order to do so, the data distributions used to train the GAN needs to be carefully chosen as follows. First, realistic canonical SIMs that are associated with a mathematical procedure for generating images need to be identified, because it allows for direct control over image properties of interest. For these canonical SIMs, statistical quantities that are medically meaningful for the particular canonical SIM need to be identified. These tasks are described next.

\begin{figure*}
     \includegraphics[width=\linewidth]{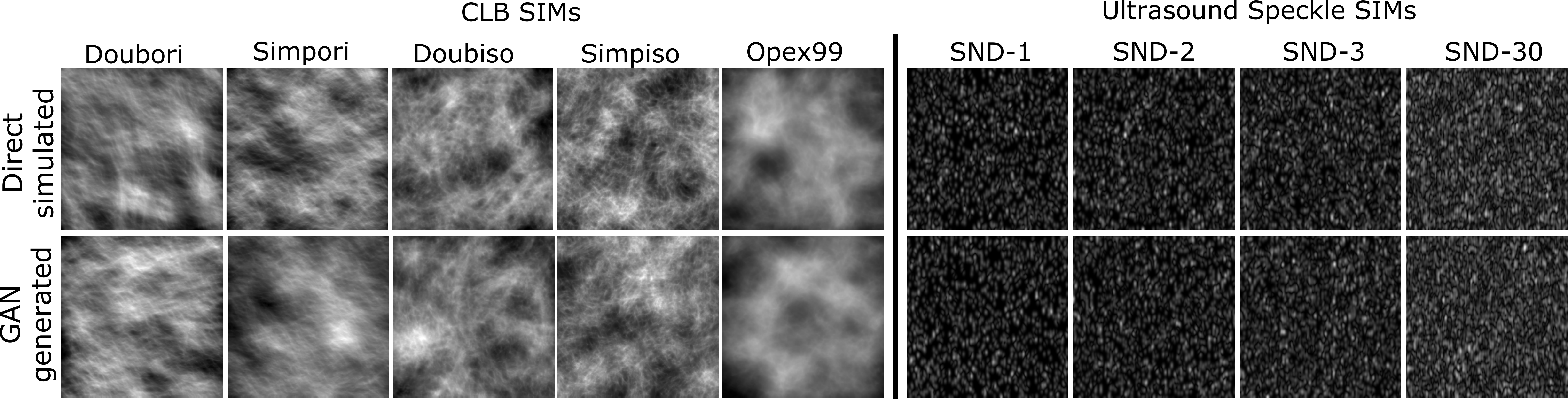}
      \caption{Images simulated from the canonical CLB and USS SIMs and images generated by the GANs trained on images from the SIMs.}
      \label{fig:real_and_fake_clb}
\end{figure*}

\begin{figure}
     \includegraphics[width=\linewidth]{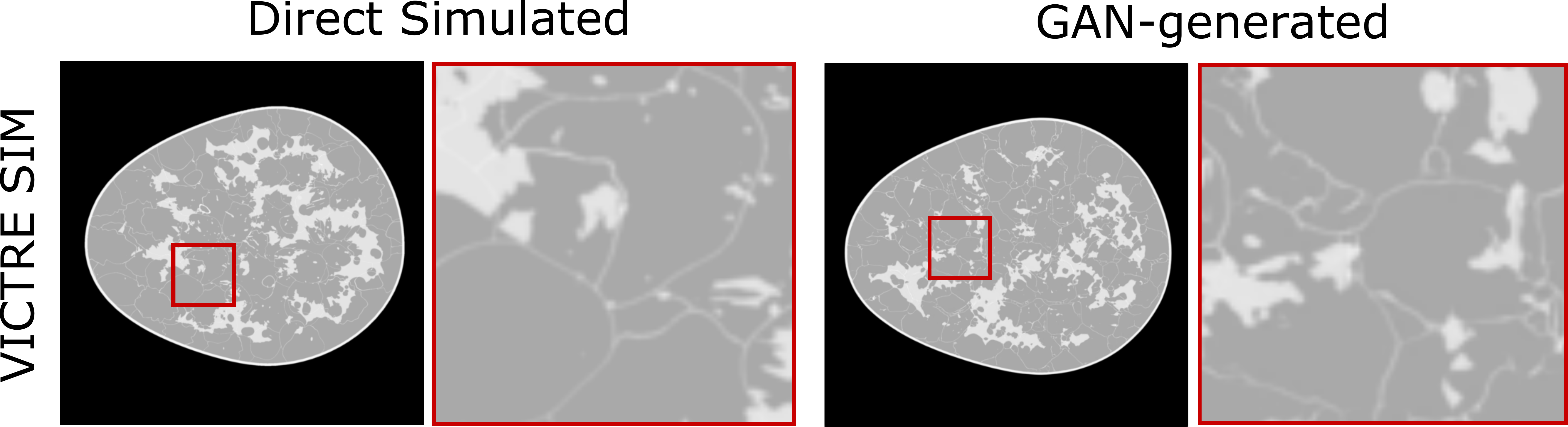}
      \caption{Images simulated from the canonical VICTRE SIM and images generated by the GAN trained on images from the SIM.}
      \label{fig:real_and_fake_victre}
\end{figure}

\subsection{Canonical stochastic image models}
Stochastic models of simulated medical images have been developed in order to approximately capture the variability in medical image distributions \cite{barrett, hmi, lumpy}. Traditionally, such stochastic image models (SIMs) have been established by developing a mathematical procedure for generating images that possess certain prescribed statistical properties. Examples of such SIMs include the lumpy background model \cite{lumpy}, the clustered lumpy background (CLB) model \cite{clb}, B-mode ultrasound speckle model \cite{insana}, among others. {Once a SIM is established, it can be used to model image statistics in virtual imaging trials \cite{vit}.} Here, the three canonical SIMs identified for the purpose of evaluating a GAN-based SIM are briefly reviewed. These SIMs are the modified clustered lumpy background model \cite{clb2}, the B-mode ultrasound speckle model \cite{insana} and the stylized 2D VICTRE breast phantom model \cite{victre}.
As compared to real medical images, simulated images from these SIMs provide the ability to examine the behavior of the GAN under a controlled setting, with several different parameter configurations of the canonical SIM.

% \section{Canonical stochastic image models for GAN assessment}\label{sec:sim}
% In this work, we wish to study the ability of medical image GANs to reproduce image statistics that are meaningful and pertinent to the medical stochastic image model under consideration, and to see how well traditional measures such as the FID correlate with these pertinent statistics. In order to do so, the data distributions used to train the GAN were carefully chosen based on the following criteria. First, realistic canonical SIMs that are associated with a mathematical procedure for generating images were identified, because it allows for direct control over image properties of interest. For these canonical SIMs, statistical quantities that are medically meaningful for the particular canonical SIM were identified. This section describes the two canonical SIMs employed in this study, namely, the modified clustered lumpy background (CLB) model \cite{clb2}, and the B-mode ultrasound speckle model \cite{insana}. Statistical texture features associated with the realism of CLB images as compared to real mammograms as assessed by radiologists \cite{clb2} were identified as the pertinent statistics of interest for the CLB model. For the B-mode ultrasound speckle model, statistical quantities associated with tissue characterization and classification \cite{insana, oelze} were identified as the meaningful statistics of interest.

\subsubsection{Modified clustered lumpy background (CLB) model}\label{sec:clb}
The CLB model was developed by Bochud et al. \cite{clb} for generating random backgrounds that resemble the image textures seen in mammography. 
In 2008, Castella, \textit{et. al} proposed variations to the original CLB model so that the images from the model better resemble realistic mammographic textures as judged by human experts \cite{clb2}. 
In addition to introducing oriented structures and long-range correlations, the authors
% The first variation was designed to favor oriented structures, similar to those occurring in real mammograms, by choosing a non-uniform distribution for the blob orientations. The second variation proposed to superimpose the CLB image with another CLB image consisting of fewer, longer blobs to model the long-range correlations in the image. 
proposed to adjust the parameters of the CLB model in order to improve the realism of the images generated. This was done by computing 17 different texture features on both the real mammographic regions of interest (ROIs) as well as images generated from the CLB model. These were used to formulate a loss function that was minimized by tuning the parameters of the CLB model.

% \subsubsection{Identification of SIM-pertinent statistics for GAN evaluation} 
% The 17 different texture features identified by Castella, \textit{et al.} mentioned above have been demonstrated to be useful for improving the medical realism of CLB images under objective and psychophysical experiments involving the judgement of radiologists \cite{clb2}. Therefore, these statistics are chosen as the statistics meaningful for assessing a GAN trained on the CLB SIMs. 
% These texture features include those derived from the per-image, gray-level intensity distribution, gray-level co-occurrence matrices (GLCMs) \cite{haralick}, primitives matrices, and the neighborhood gray tone difference matrix (NGTDM) \cite{ngdtm}. The details of computation of these texture features are described in Section \ref{sec:num_methods}.

\subsubsection{B-Mode Ultrasound Speckle (USS)}

% \begin{align}\label{eqn:us-envelope}
%     E = \sum_{i=1}^N a_i,
% \end{align}
% where $a_i = |a_i|\exp(j\phi_i)$ with phases $\phi_i$ that are uniformly distributed on $[0,2\pi]$ and are statistically independent. 

B-mode ultrasound speckle (USS) can be viewed as a random phasor sum of complex signals \cite{insana}. The received complex signal $E$ is a radio frequency voltage output from an ultrasound transducer and can be modeled as the sum of $N$ complex signals with phases statistically independent uniformly distributed on $[0,2\pi]$ \cite{insana}. The quantity $N$ is the number of scatterers per resolution cell or equivalently the scatterers per number density (SND) times the resolution cell size. The resolution cell size is defined as the axial resolution ($\rm AR$) times the lateral resolution ($\rm LR$), given in Ref. \cite{iaea_handbook},
where the parameters are the frequency of the carrier $f_c$, the wave velocity $v$, the ratio between the focal distance and the length of the aperture (called the $f$-number) and the number of cycles within the full width half maximum in the spatial direction (FWHM) $N_c$ . The USS SIM is modeled using the method proposed in Ref. \cite{insana} where the standard deviations of the 2-D Gaussian PSF are determined by the AR and LR.

% \begin{align}\label{eqn:axial_resolution}
%     {\rm AR} = \frac{N_c \lambda}{2}, \quad\quad {\rm LR} = \frac{\lambda F}{L},
% \end{align}
% where $\lambda$ is the wavelength of the carrier, $F$ is the focal distance, $L$ is the length of the aperture, and $N_c$ is the number of cycles within the full width half maximum in the spatial direction (FWHM) which is a number between 1.5 and 2 \cite{iaea_handbook}. 
% The lateral resolution is given as

% \begin{align}\label{eqn:lateral_resolution}
%     LR = \frac{\lambda F}{L},
% \end{align}

% where $F$ is the focal distance and $L$ is the length of the aperture. The ratio $F/L$ is the $f$ number of a transducer and is usually a number between 2 and 6 \cite{iaea_handbook}.

If $N$ is large, the resulting USS follows Gaussian statistics and is called fully developed speckle. In this case, the envelope $|E|$ follows a Rayleigh distribution and thus the intensity $I = |E|^2$ follows an exponential distribution.
% Because the envelope follows a Rayleigh distribution, the point-wise signal to noise ratio ($\rm SNR_0$) is always 1.91. 
If $N$ is small then the resulting USS is called non-Gaussian speckle and its statistical properties are determined by $N$ \cite{insana}. 

% The USS SIM can be modeled by distributing scatterers over the image area using a Poisson point process with the mean parameter equal to the SND times the image area. The $(x,y)$ positions are then chosen using a uniform distribution \cite{insana}. For theoretical ultrasound speckle, the scatterers are dimensionless \cite{iaea_handbook} and are thus represented by points in space. Assuming the propagation of the carrier wave is in the $x$ direction, then for each scatterer the complex phasor applied is $e^{\frac{j 2 \pi x}{\lambda}}$. The resulting scatterers are then convolved with a 2-D Gaussian which simulates the system point spread function (PSF)\cite{insana}. The standard deviations for the 2-D Gaussian PSF are determined by the axial and lateral resolutions. % \autoref{eqn:axial_resolution}.

\subsubsection{The stylized 2D VICTRE (S2V) breast phantom model}
% The Virtual Imaging Clinical Trials for Regulatory Evaluation (VICTRE) project within the USS Food and Drug Administration (FDA) has provided computational tools for simulating the entire imaging pipeline for digital breast tomosynthesis. The VICTRE software tool enables the generation of random anthropomorphic numerical phantoms of the human female breast with realistic anatomical structures and variability in patient-specific characteristics such as the breast type, shape, size, density and granularity. Each numerical breast phantom (NBP) is a 3D voxelized map of several tissue types, such as fat, skin, glandular tissue, artery, vein, and so on. By appropriately assigning physical attributes and material coefficients to the different tissue types, the VICTRE simulation tool can be customized to create canonical SOMs corresponding to different imaging modalities. 

The US Food and Drug Administration's (FDA) Virtual Imaging Clinical Trials for Regulatory Evaluation (VICTRE) initiative has produced a set of software tools for simulating random anthropomorphic phantoms of the human female breast \cite{victre}. 
These numerical breast phantoms (NBPs) are three dimensional (3D) voxelized maps, where a voxel value denotes the tissue type from one out of the following 10 tissues: fat, glandular tissue, skin, artery, vein, muscle, ligament, nipple and terminal duct lobular unit. Controlling the patient-specific input parameters such as breast type, size, shape, granularity and density, and setting the random seed number enables the generation of large ensembles of stochastic NBPs with realistic variation in breast anatomy, shape and fat-to-glandular tissue ratio. The VICTRE model is thus a general stochastic object model (SOM) that can be specialized to different imaging modalities by assigning the appropriate physical coefficients. In particular, by assigning X-ray linear attenuation coefficients to the various tissues in the NBPs and extracting 2D slices from the 3D phantom, a SIM can be obtained. 
The VICTRE software creates NBPs that correspond to four breast types identified by the American College of Radiology’s (ACR) Breast Imaging Reporting and Data System (BI-RADS) \cite{birads} and are distinguished by the amounts of fat and glandular tissue.

\section{Numerical studies}\label{sec:num_methods}

\subsection{SIM training data and GAN training}
\subsubsection{The CLB model}
The following four parameter configurations of the modified CLB model that were shown to produce realistic simulated mammographic images under radiologists' assessment \cite{clb2} were used in this study -- (1) \textit{doubiso}, a double-layered CLB model with isotropically oriented clusters, (2) \textit{simpiso}, a single-layered CLB model with isotropically oriented clusters, (3) \textit{doubori}, a double-layered CLB model with anisotropically oriented clusters, and (4) \textit{simpori}, a single-layered CLB model with anisotropically oriented clusters. Additionally, images from the original CLB model \textit{opex99}, proposed by Bochud \textit{et al.} \cite{clb} were employed. The gray levels and pixel value range were set in accordance with Castella \textit{et al.}  \cite{clb2}. For each of the five canonical SIMs, a GAN was trained on a dataset of 100,000 256$\times$256 images from the SIM.

As discussed in the Introduction, medical image distributions are typically mixed distributions consisting of multiple classes or modes. In order to illustrate the effect of mixing distributions on the identified SIM-pertinent measures,
a stylized emulation of data coming from two different imaging sites or clinical systems having different resolution properties was constructed. Accordingly, one of the classes consisted of \textit{doubiso} images as are described above. The other class consisted of \textit{doubiso} images that were first degraded by use of a Gaussian blur followed by low-pass filter $\mathcal{H}_{\rm LPF}(\cdot)$ with cutoff at half the image bandwidth. Two such multi-class datasets were constructed, one having a 50\textbackslash50\% split and the other having a 95\textbackslash5\% split between the regular and degraded image classes. These two datasets will henceforth be referred to as the \textit{doubiso} 50-50 and \textit{doubiso} 95-5 datasets respectively.

\subsubsection{B-mode Ultrasound Speckle Model} 
% All parameter configurations for USS were simulated using the method described in Wear \textit{et al.} \cite{insana}.
% Section \ref{sec:bkd}D. 
The parameter configurations chosen for the USS SIMs are follows. All images were $256 \times 256$ pixels in size {with each pixel corresponding to a $100 \mu \text{m} \times 100 \mu \text{m}$ square.} The velocity of the wave was set to $v = 1556 \text{m/s}$, the frequency $f_c$ was set to 3.5 MHz, the number of cycles within the FWHM was set to $N_c=2$, the $f$-number for the $y$ direction was set to $2$ and the $f$-number in the $z$ direction was set to $3$. The ultrasound wave was assumed to be propagating in the $x$ direction. 
The SND parameter was varied to create four canonical USS SIM datasets, corresponding to SND values of 1, 2, 3 and 30 $\text{mm}^{-3}$ respectively. The first three values were chosen because they fall in the range of SND values that can be accurately estimated from the image \cite{insana}, which is not the case for the \textit{SND-30} SIM that represents a fully developed speckle \cite{iaea_handbook}. These four SIMs will henceforth be called (1) \textit{SND-1}, (2) \textit{SND-2}, (3) \textit{SND-3} and (4) \textit{SND-30} respectively. 
%All other parameters for the USS model were set within the range of acceptable values provided in the literature \cite{iaea_handbook}.

Additionally, similar to the CLB case, two multi-class datasets were considered, where \textit{SND-2} and \textit{SND-3} were (1) distributed with a 50\% - 50\% split and (2) were distributed with a 95\% - 5\% split. Henceforth these datasets will be called \textit{USS Mixed 50-50} and \textit{USS Mixed 95-5}.

% In total, twelve GANs were trained, each with 100,000 images from the above SIMs. Before training the GANs using these images, the top $1\%$ of the ensemble pixel histogram values for each unique USS dataset was clipped and the maximum value of the clipped ensemble was set to 255. All USS are grayscale unsigned 8-bit images.

For each of the above described SIMs, a GAN was trained using 100,000 images from the SIM. Before training, each ensemble of training images was converted to an unsigned, 8-bit grayscale where 255 corresponds to the top 1\% pixel value in the ensemble.

%Before training the GANs using these images, the top $1\%$ of the ensemble pixel histogram values for each unique USS dataset was clipped and the maximum value of the clipped ensemble was set to 255. All USS images were grayscale unsigned 8-bit images.

\subsubsection{The S2V model}
The S2V was obtained from the 3D VICTRE NBP SOM described in \autoref{sec:bkd} as follows. First, a dataset of 1000 3D NBPs was generated using the VICTRE tool \cite{victre}. Next, linear attenuation coefficients in $\rm cm ^{-1}$ for X-rays of energy 30 keV were assigned to the pixels corresponding to each of the tissue types. These values were either directly obtained from literature, or calculated using the mass attenuation coefficient and material density values obtained from literature \cite{chen, nist, tomal}. Coronal slices were extracted from a central region of an NBP that ranges from 40\% through 70\% of the distance from the outermost coronal plane to the innermost coronal plane. This was done to avoid extracting slices too close to the chest wall or the nipple. A spacing of 50 pixels was maintained between two slices consecutively extracted from the same NBP. The extracted slices were then downsampled to an image dimension of 512$\times$512, which corresponds to the length scale of 0.4 $\mu$m per pixel.
The described procedure generated a 2D dataset of 130,000 slices, which was used for training a GAN.

% \input{images/tab_attenuation_values}

% \subsection{GAN architecture and training}
{StyleGAN2, proposed by Karras \textit{et al.} \cite{stylegan2} was employed as the GAN in all the studies described in this work. All the default parameters and configurations of the StyleGAN2 architecture including the latent space dimensionality were kept the same as the the original code base, except for the number of channels in the output image, which was set to 1. The networks were trained using Tensorflow 1.14/Python \cite{tensorflow} on an Intel Xeon Gold 5218 CPU and two Nvidia Quadro RTX 8000 GPUs.}

\subsection{Identification and computation of SIM-pertinent evaluation measures}
A GAN may learn different types of image statistics to different levels of correctness. Hence, it is important to evaluate GANs using measures based on those statistics that are meaningful and pertinent to the SIM considered. In this study, such SIM-pertinent evaluation measures are based on statistics that either have been deemed important for assessing the realism of the canonical SIM images by human experts, or are known to be related to biomarkers important for a particular diagnostic task. These statistics are computed from both the ``direct-simulated" images, i.e. images directly simulated from the canonical SIM, as well as the GAN-generated images. 

\subsubsection{The CLB model}
The 17 different texture features identified by Castella, \textit{et al.} mentioned in Section \ref{sec:bkd} have been demonstrated to be useful for improving the medical realism of CLB images under objective and psychophysical experiments involving the judgement of radiologists \cite{clb2}. Therefore, these statistics were chosen as the statistics meaningful for assessing a GAN trained on the CLB SIMs. 
These texture features include those derived from the per-image, gray-level intensity distribution, gray-level co-occurrence matrices (GLCMs) \cite{haralick}, primitives matrices (GLRM), and the neighborhood gray tone difference matrix (NGTDM) \cite{ngdtm}.

For each of the five CLB model types in Section \ref{sec:num_methods}A, as well as the two multi-class CLB SIMs, the following 17 texture features described by Castella, \textit{et al.} \cite{clb2} were computed from each image of the evaluation datasets.
Mean, standard deviation, skewness and kurtosis were derived from the per-image gray-level intensity distribution. The texture features energy, entropy, maximum, contrast and homogeneity were computed from the GLCMs.
Four features were derived from the primitives matrices (GLRMs), namely, the short primitive emphasis (SPE), long primitive emphasis (LPE), gray level uniformity (GLU), and primitive length uniformity (PLU). The four features derived from the NGTDM \cite{ngdtm} were coarseness, contrast, complexity and strength. Various parameter values required for the computation of the texture features, such as the number of gray levels, two-point distances and angles were fixed to the values used in Castella, \textit{et al.} \cite{clb2}. The resulting feature data were then used for further analysis in order to summarize trends. Two types of analyses were conducted on the feature data. The first computed an empirical estimation of the JS divergence between the joint texture-feature distributions by utilizing the feature data \cite{empirical_kl}. The second plotted the joint empirical PDF over the first two principal components of texture features. The texture features used for this computation were selected as follows. First, principal component analysis (PCA) was conducted for each of the three spatial texture feature families, namely -- the GLCM, GLRM and NGTDM feature families. Next, the first two principal components were selected, and an empirical PDF over these two components was computed. The empirical PDFs that give the highest discrepancy between the direct-simulated and GAN-generated distributions in terms of the total variation (TV) distance were plotted.

\subsubsection{B-mode Ultrasound Speckle Model} 
% Wagner \textit{et al.} \cite{wagner_auto} proposed the autocorrelation and the corresponding Fourier power spectrum of E as a means of deducing more information about the object structure in USS images.\\
Previous studies have shown that the intensity signal-to-noise ratio (SNR) of USS images is associated with the envelope statistics \cite{oelze}. In regions of the body such as the liver and the breast, the envelope statistics have previously been successfully used for tissue characterization \cite{oelze}. Therefore, it was chosen as the SIM-pertinent statistic, though this preliminary study does not associate a given speckle model with a tissue type. 

The PDF of the $\rm SNR^2$ estimate of USS speckle can be modeled as a Gaussian distribution centered around the true $\rm SNR^2$. If the scatterers per resolution cell $N$ follows a Poisson distribution, then one can estimate $N$ using $\rm SNR^2$. The SNR and $N$ estimate called $\hat{N}$ are defined as:

\begin{align}\label{eqn:SNR_2}
    {\rm SNR} = \frac{\mu_I}{\sigma_I}, \text{ } \hat{N} = \frac{\rm SNR^2}{1-\rm SNR^2},
\end{align}

where $\mu_I$ and $\sigma_I$ are the mean and standard deviation of the intensity. The SNR and $\hat{N}$ were computed on a per-image basis for both the direct-simulated and GAN-generated images using empirically estimated $\mu_I$ and $\sigma_I$ from each image in the test dataset. The JS divergence was used as a measure to summarize the discrepancy between the $\rm SNR^2$ PDFs of the direct-simulated and GAN-generated images.

% The SNR is defined as follows \cite{insana}:

% \begin{align}\label{eqn:SNR_2}
%     \rm{SNR} = \frac{\mu_I}{\sigma_I},
% \end{align}
 
% where $\mu_I$ and $\sigma_I$ are the mean and standard deviation of the intensity. The probability density function of the $\rm SNR^2$ estimate of USS speckle can be modeled as a Gaussian distribution centered around the true $\rm SNR^2$. If the scatterers per number density $N$ follows a Poisson distribution, then one can estimate $N$ using $\rm SNR^2$ as \cite{insana}

% \begin{align}\label{eqn:N_estimate}
%     \hat{N} = \frac{SNR^2}{1-SNR^2}.
% \end{align}

\subsubsection{The S2V model} 
Human female breasts can be categorized into four different types based on the relative amount of fat and glandular tissue \cite{birads}. It is known that the amount of fat compared to the glandular tissue is an important factor impacting the risk of developing breast cancer, and the effectiveness of screening tests such as mammography in detecting breast masses \cite{wolfe1, fgr, birads}. Fat and glandular tissue have different linear attenuation coefficients \cite{chen, nist, tomal}. Therefore, the ratio of fat-to-glandular tissue was chosen as the SIM-pertinent measure for evaluating the GAN trained on the S2V SIM. For the idealized S2V SIM described in \autoref{sec:num_methods}A, the ratio $\rho_{F:G}$ of the amount of fat-to-glandular tissue in a thin coronal slice of an NBP can be computed by first calculating the number of pixels $F$ and $G$ relative to the total image pixels having linear attenuation coefficient values close to that of fat and glandular tissue respectively, and then computing their ratio $\rho_{F:G} = F / G$. Because the linear attenuation coefficient value of fat and glandular tissue are far enough to not confound a simple thresholding-based segmentation scheme, the values of $F, G$ and $\rho_{F:G}$ can be estimated accurately both for the direct-simulated and GAN-generated images. 
Using this procedure, $\rho_{F:G}$ was estimated on a per-image basis for both the direct-simulated and GAN-generated images. The empirical PDFs of $\log \rho_{F:G}$ computed from both the direct-simulated and GAN-generated images were plotted, and the JS divergence between the two PDFs was computed.

Apart from the above-described SIM-pertinent measures, basic ensemble statistics, such as the histogram of gray level values and the empirical image autocorrelation were computed from direct-simulated and GAN-generated images from all the SIMs in order to assess the ability of the GAN to learn these statistics accurately. As described in Bochud \textit{et al.} \cite{clb}, a Papoulis window was used in order to overcome boundary artifacts in the computation of the autocorrelation.
The FID score between a direct-simulated and a GAN-generated test dataset, as well as two i.i.d. direct-simulated datasets was computed. The latter serves as a heuristic noise floor for the FID score for the particular SIM. A pre-trained InceptionV3 network \cite{inception} was employed for this purpose. All the evaluation measures were computed using 10,000 direct-simulated and GAN-generated images. Other test dataset sizes were examined, and the computed metrics were found to be qualitatively no different.

\begin{figure}
    \includegraphics[width=\linewidth]{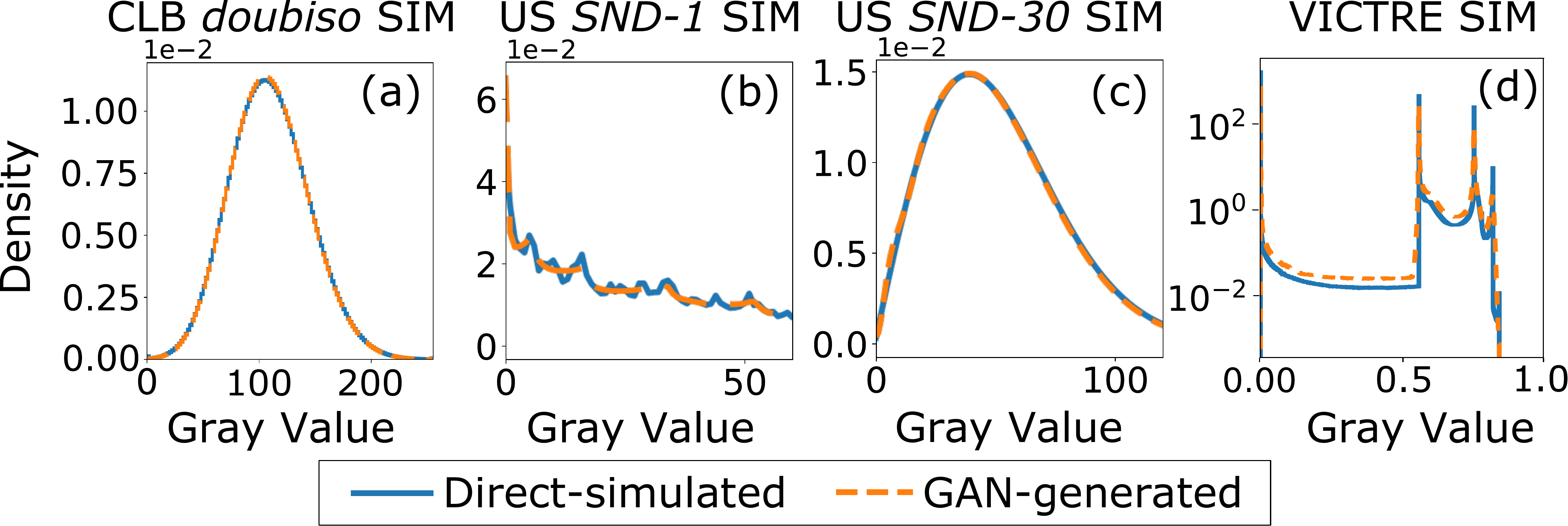}
    \caption{Sample empirical gray level PDFs of direct simulated and GAN-generated images for the three types of SIMs.}
    \label{fig:intensity_histogram}
\end{figure}

\begin{figure}
\includegraphics[width=\linewidth]{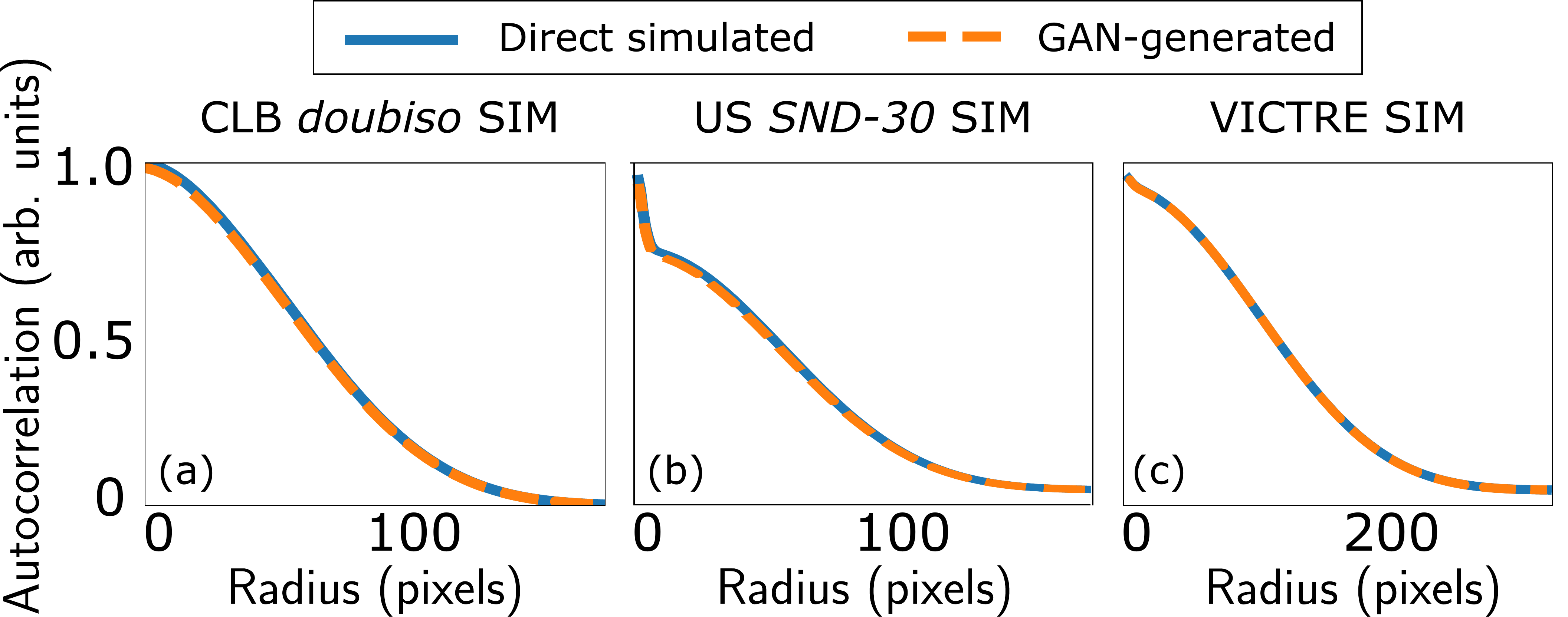}
\vspace{1pt}
\caption{Sample radial profiles of autocorrelation of direct simulated and GAN-generated images for the three types of SIMs.}
\label{fig:autocorrelation}
\vspace{-10pt}
\end{figure}

\section{Results}\label{sec:results}
% \subsection{Impact of training on the traditional and meaningful measures}

% \subsection{Basic First and second-order Statistics}

This section is organized as follows. Section \ref{sec:results}A qualitatively describes the images generated by the GAN. Section \ref{sec:results}B describes the basic ensemble statistics learned by the GAN, such as the intensity histogram and the image autocorrelation. Section \ref{sec:results}C describes and compares the FID score and the identified meaningful measures based on their ability to assess the fidelity of the trained GAN. Finally, Section \ref{sec:results}D compares the ability of the FID score and the identified measures to assess multi-modal SIMs. 

\begin{figure}
     \includegraphics[width=\linewidth]{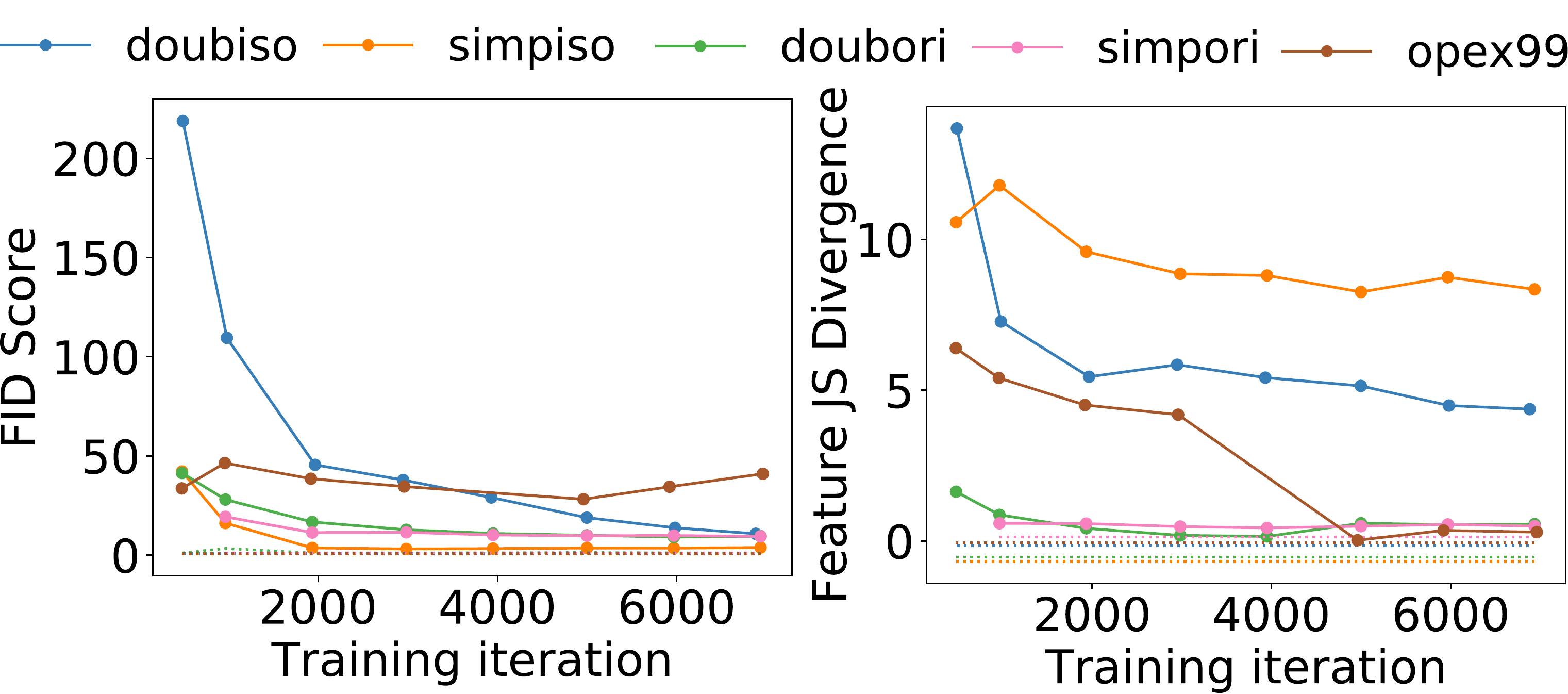}
      \caption{FID and empirical feature JS divergence measures between the real and GAN-generated distrbutions for \textit{opex99}, \textit{simpiso}, and \textit{doubiso} models. The dotted lines represent the value of the measures between two direct-simulated datasets.}
      \label{fig:fid_js_singleclass}
\end{figure}

\begin{figure}
     \includegraphics[width=\linewidth]{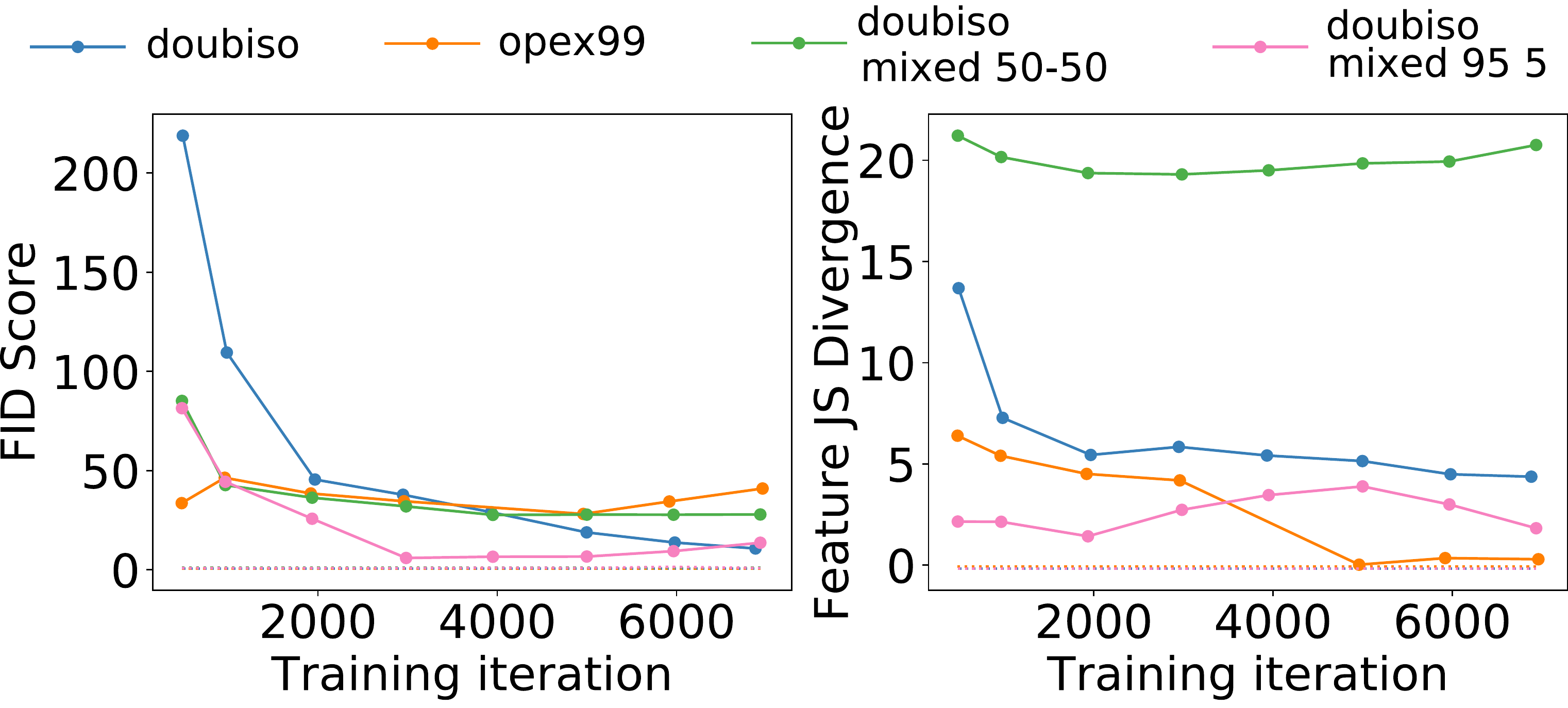}
      \caption{FID and empirical feature JS divergence between the real and GAN-generated distrbutions for \textit{opex99}, \textit{doubiso}, and the two \textit{doubiso-mixed} models. The dotted lines represent the value of the measures between two direct-simulated datasets.}
      \vspace{-10pt}
      \label{fig:fid_js_multiclass}
\end{figure}

\begin{figure*}
     \includegraphics[width=\linewidth]{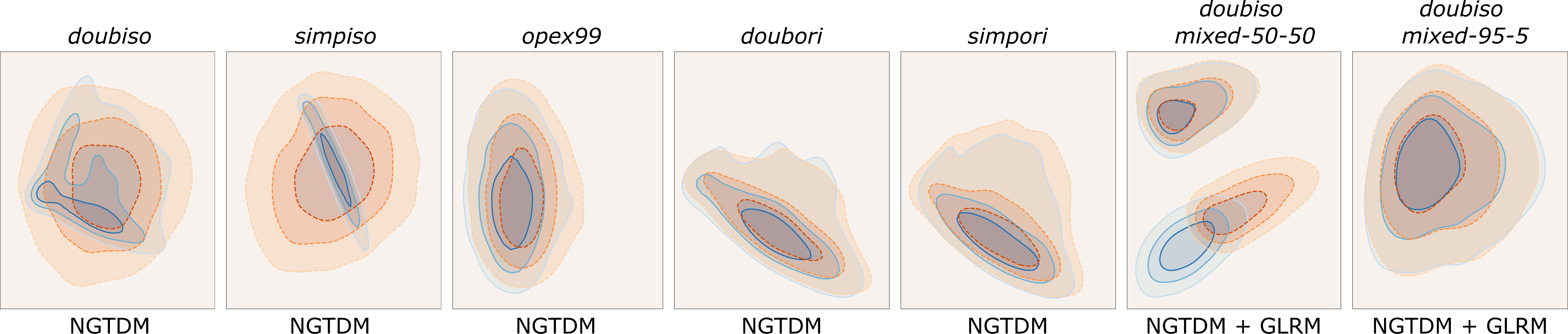}
      \caption{Empirical PDF over the first two principal components of the CLB feature data. The selected texture feature family for each of the models is shown below each plot. The blue and the orange contour plots denote the direct-simulated and GAN-generated distributions respectively.}
      \label{fig:clb_worst_components}
\end{figure*}

\begin{figure}
     \includegraphics[width=\linewidth]{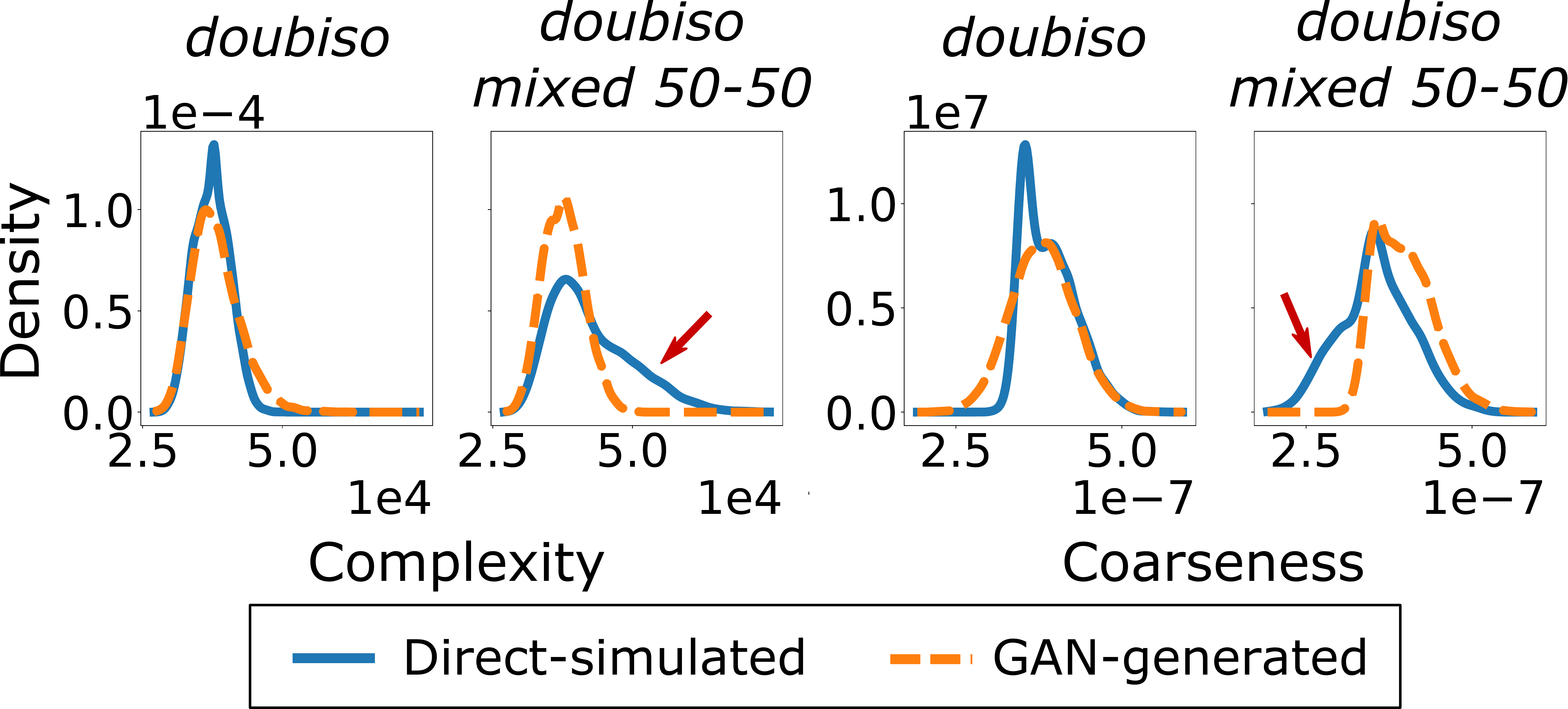}
      \caption{Distributions of per-image NGTDM Complexity and Coarseness features learned by the GAN for the \textit{doubiso} and \textit{doubiso mixed 50-50} SIMs. The red arrows point to the parts of the distribution corresponding to the degraded class that are completely ignored by the GAN.}
      \label{fig:clb_worst_components_individual}
\end{figure}

\begin{figure}
    \includegraphics[width=\linewidth]{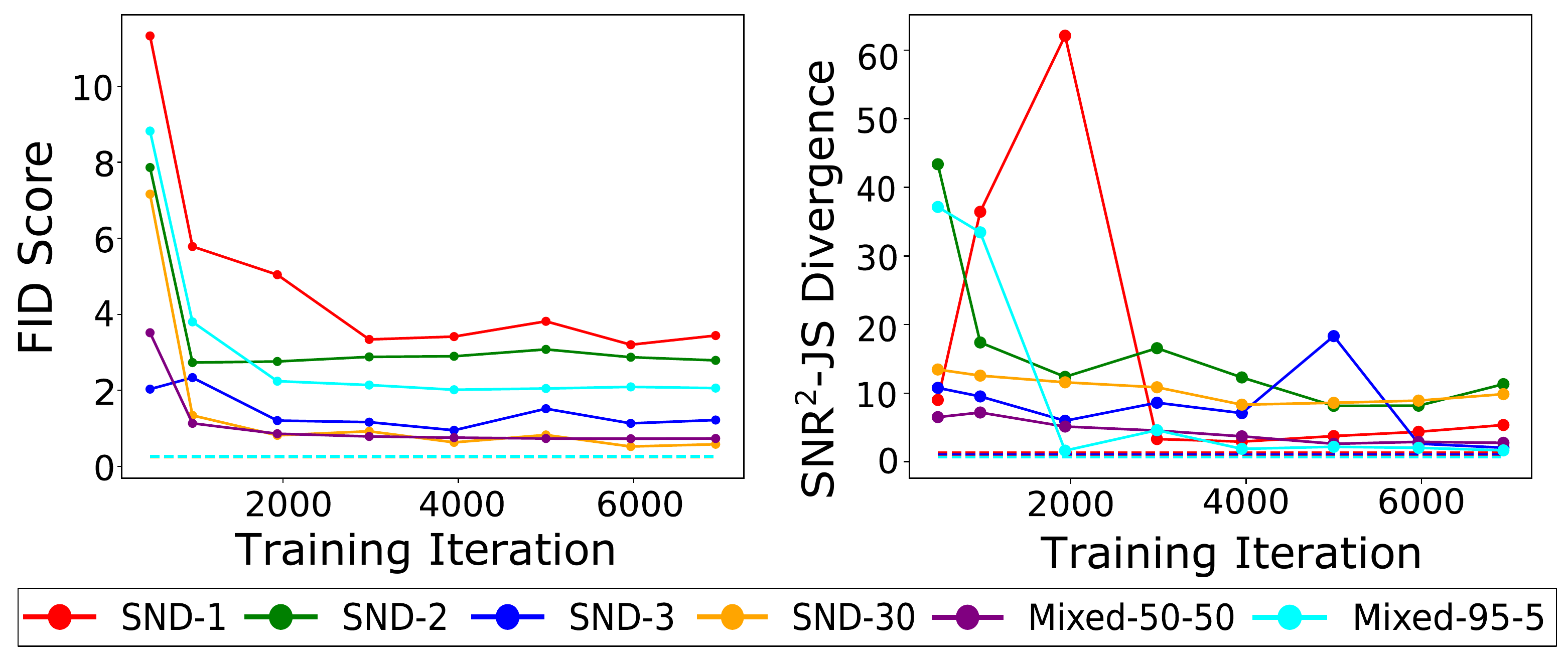}
    \caption{FID and $\rm SNR^2$-JS divergence between the real and GAN-generated distributions for \textit{SND-1}, \textit{SND-2}, \textit{SND-3}, \textit{SND-30}, \textit{USS Mixed 50-50} and \textit{USS Mixed 95-5}. The dotted lines represent the value of the measures between two direct-simulated datasets.}
    \label{fig:US_FID_KL}
    \vspace{-10pt}
\end{figure}

\begin{figure*}
     \includegraphics[width=\linewidth]{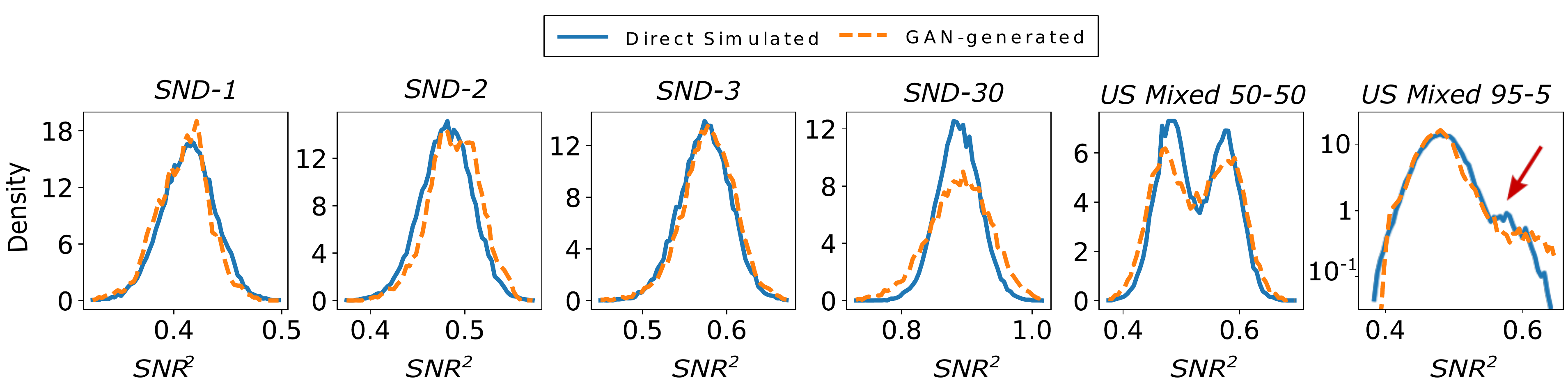}
      \caption{Estimated $\rm SNR^2$ PDFs of both direct-simulated and GAN-generated images for \textit{SND-1}, \textit{SND-2}, \textit{SND-3}, \textit{SND-30}, \textit{USS Mixed 50-50} and \textit{USS Mixed 95-5}. Although the direct-simulated and GAN-generated distributions tend to match well, occasionally this is not the case as can be seen in for \textit{SND-30} and \textit{USS Mixed 50-50}. Note that the \textit{USS Mixed 95-5} $\rm SNR^2$ PDF has the density in log scale with the red arrow pointing to the distribution of the \textit{SND-3} class having 5\% prevalence.}
      \label{fig:SNR_2_us}
\end{figure*}

\subsection{Qualitative assessment of images generated by the GAN}
Figures \ref{fig:real_and_fake_clb} and \ref{fig:real_and_fake_victre} show the images generated by the trained GANs alongside direct-simulated images from the training dataset for the single-class CLB, USS and S2V models. It can be seen that there is obvious visual similarity between the direct-simulated and the GAN-generated images. Note that this is even true for the zoomed-in images of the S2V model shown in Fig. \ref{fig:real_and_fake_victre}. One important thing to note, however, is that some of the ligaments in the GAN-generated images appear broken at certain locations, which is not the case for the direct-simulated images.

\subsection{Basic ensemble statistics learned by GANs}
Figure \ref{fig:intensity_histogram} shows the ensemble empirical PDF of pixel gray levels for the CLB \textit{doubiso} SIM, the USS \textit{SND-1} and the \textit{SND-30} SIMs, and the S2V SIM, computed from both the direct-simulated and GAN-generated images. A close match between these empirical PDFs indicates that the GAN is able to reproduce first-order statistics. The GAN performs similarly for the other CLB SIMs, which have gray-level distributions similar to the ones shown in Fig. \ref{fig:intensity_histogram}a. It can be seen that for USS \textit{SND-30} SIM, which represents a fully developed speckle, the GAN reliably reproduces the expected Rayleigh distribution of grayscale values. For the USS \textit{SND-1} SIM, this distribution is far from Rayleigh both for the direct-simulated and GAN-generated images, yet the GAN recovers this distribution successfully. The pixel-value distributions corresponding to USS \textit{SND-2} and \textit{SND-3} SIMs appear intermediate between the ones shown in Fig. \ref{fig:intensity_histogram}b and c. 

Fig. \ref{fig:autocorrelation} shows the radial profile of the image autocorrelation computed using the direct-simulated and GAN-generated images for the CLB \textit{doubiso}, USS \textit{SND-1} and S2V SIMs. It can be seen that the GAN was successful in recovering this particular second-order statistic. Similar results were obtained for the other CLB and USS SIMs considered.

\subsection{SIM-pertinent measures learned by GANs}

\subsubsection{CLB Model}
Figure \ref{fig:fid_js_singleclass} shows the FID as well as the texture feature JS divergence between the direct-simulated and GAN-generated distributions as a function of training iteration. In Fig. \ref{fig:fid_js_multiclass}, the FID scores and the feature JS divergences for the \textit{doubiso mixed 50-50} and \textit{doubiso mixed 95-5} datasets are shown along with those for the single class \textit{doubiso} and \textit{opex99} models. It can be seen that as the training progressed, both the FID as well as the empirical feature JS divergence converged for most of the SIMs considered. However, in some cases, these measures either diverged or varied erratically as the training progressed. 
Furthermore, the high value of the feature JS divergence for the GAN trained on the \textit{doubiso mixed 50-50} model suggests that the GAN was not able to reproduce the meaningful feature statistics as well as the GAN trained on the single class dataset. On the other hand, the FID plot in Fig. \ref{fig:fid_js_multiclass} shows comparable FID scores for the various SIMs and does not predict the same trend as the feature JS divergence plots. This suggests that for this specific example, the FID score could be blind to telling if multiple modes in the distribution are learned correctly.

{These findings were further investigated using the principal components of the data from the texture feature family that was learnt the least accurately by the GAN. The procedure for computing these components was described earlier in Section \ref{sec:num_methods}B.} Figure  \ref{fig:clb_worst_components} plots this joint empirical PDF for the direct-simulated and GAN-generated images. Note that these texture features are computed on a per-image basis. For most of the CLB SIMs, obvious dissimilarities between the original and learned distributions can be seen. These dissimilarities correlate well with the feature JS divergence values shown in Figures \ref{fig:fid_js_singleclass} and \ref{fig:fid_js_multiclass}, but not with the corresponding FID values. For the \textit{doubiso mixed 50-50} SIM, it can be seen that the GAN failed to correctly learn the distribution of principal NGTDM and GLRM texture components for one of the classes. On further investigation and comparison with the individual texture distributions for the \textit{doubiso} SIM, it was revealed that the GAN failed to learn the per-image NGTDM coarseness and complexity distributions of the images from the degraded class, as shown in Fig. \ref{fig:clb_worst_components_individual}. This was despite the GAN being able to learn ensemble measures such as the FID and basic first- and second-order statistics well.

% For the CLB SIMs, it can be seen that the empirical JS divergence approaches the noise floor, but still converges to a value higher than the noise floor for \textit{simpiso} and \textit{doubiso} image types. This phenomenon is more evident in the case of \textit{simpiso} than \textit{doubiso}, indicating that the GAN trained on \textit{doubiso} trains better with respect to the feature distribution JS divergence. This, however, is in contrast to the findings from the FID scores shown in Fig. \ref{fig:fid_kl_singleclass}, which shows that the GAN trained on the \textit{simpiso} distribution had a lower FID score than the GAN trained on the \textit{doubiso} and \textit{opex99} distributions. %Discuss PCA plot here

\begin{table}[h!]
\centering
\resizebox{0.5\textwidth}{!}{
\begin{tabular}{lcccccccc}
\toprule
 & \multicolumn{2}{c}{\textit{SND-1}} & \multicolumn{2}{c}{\textit{SND-2}} & \multicolumn{2}{c}{\textit{SND-3}} & \multicolumn{2}{c}{\textit{SND-30}}\\
\cmidrule(lr){2-3} \cmidrule(lr){4-5} \cmidrule(lr){6-7} \cmidrule(lr){8-9}
 & D.S. & G.G. & D.S. & G.G. & D.S. & G.G. & D.S. & G.G.\\
\midrule
$\mu$ & 0.41 & 0.41 & 0.48 & 0.49 & 0.58 & 0.58 & 0.89 & 0.89 \\
\midrule
$\sigma$ & 0.024 & 0.024 & 0.027 & 0.028 & 0.03 & 0.03 & 0.032 & 0.047 \\
\midrule
\textit{MSE} ($10^{-4}$) & 2.9 & 20.9 & 3.0 & 24.4 & 2.6 & 3.3 & 1.3 & 2.6 \\
\midrule
$\hat{N}$ & 0.71 & 0.70 & 0.94 & 0.97 & 1.37 & 1.39 & 7.54 & 9.78 \\
\bottomrule
\end{tabular}}
\caption{A table showing the mean $\mu$ and standard deviation $\sigma$ of the gaussian fit curve for both direct-simulated and GAN-generated $\rm SNR^2$ distributions, the mean squared error (MSE) between the gaussian fit and their respective $\rm SNR^2$ distributions and the mean scatterers per resolution cell estimate $\hat{N}$ of both direct-simulated (D.S.) and GAN-generated (G.G.) images.}
\label{tab:us_gauss_fit}
\vspace{-10pt}
\end{table}

\subsubsection{B-Mode Ultrasound Speckle Model} 
The empirical JS divergence between the estimated $\rm SNR^2$ PDFs computed from the direct-simulated and GAN-generated USS images (henceforth refered to as the $\rm SNR^2$-JS divergence) is shown in Fig. \ref{fig:US_FID_KL} alongside the FID score computed between the direct-simulated and the GAN-generated images. 
Although the $\rm SNR^2$-JS divergence approaches the noise floor and converges for most SIMs, it behaves erratically for a few SIMs, even as the FID score for the corresponding SIM converges.

In Fig. \ref{fig:SNR_2_us} the estimated $\rm SNR^2$ PDFs are plotted for both direct-simulated and GAN generated USS images. As can be seen the GAN generated images tend to give $\rm SNR^2$ distributions that somewhat match those of the direct-simulated images for the \textit{SND-1}, \textit{SND-2} and \textit{SND-3} SIMs. Since the $\rm SNR^2$ is theoretically expected to be distributed as a Gaussian for these SIMs \cite{wagner_auto}, 
% the Gaussian behavior of each $\rm SNR^2$ distribution was investigated quantitatively. 
each distribution of direct-simulated and GAN-generated images was fit to a Gaussian. In Table \ref{tab:us_gauss_fit} the mean and standard deviation of the best fit Gaussian distribution are shown in the first two rows while the third row shows the mean squared error between a given $\rm SNR^2$ distribution and its Gaussian fit.
% , as measure of the Gaussianity of the $\rm SNR^2$ distributions. 
The results for the mean and standard deviation of the Gaussian fit distributions confirm our visual inspection. The mean values were near perfect matches and so are the standard deviations with the exception of \textit{SND-30}.
However, the \textit{MSE} between the GAN-generated empirical $\rm SNR^2$ PDFs and their Gaussian fits was larger than the \textit{MSE} between the direct-simulated empirical $\rm SNR^2$ PDFs and their Gaussian fits.
% Although, the \textit{MSE} for all for each GAN-generated $\rm SNR^2$ distribution when compared with its best Gaussian fit, was larger than for the direct-simulated images the 
% Nevertheless, these \textit{MSE} values still remain very small indicating that all of the distributions are essentially gaussian in nature. 
Finally, the mean estimate of scatterers per resolution cell $\hat{N}$ computed from GAN-generated images was close to that computed from the direct-simulated images for all USS SIMs except for \textit{SND-30}. This is expected since the $\rm SNR^2$ distributions do not match well for the \textit{SND-30} SIM. 
% Recall from Section \ref{sec:bkd} that the scatterers per resolution cell is the scatterers per number density (SND) times the resolution cell size.

% These results are further reaffirmed by Fig. \ref{fig:US_FID_KL} where the JS divergence shows that upon completion of training the difference between direct-simulated and GAN-generated $\rm SNR^2$ distributions for all three datasets are ranked in the same order as our visual comparison between distributions.

In Fig. \ref{fig:US_FID_KL}, the FID scores and the $\rm SNR^2$-JS divergences can be seen for \textit{USS Mixed 50-50} and \textit{USS Mixed 95-5} SIMs. As the training progresses, both the measures seem to converge in a similar fashion to most of the single class SIMs. Interestingly, the \textit{USS Mixed 95-5} SIM has one of the higher FID scores while also having the lowest $\rm SNR^2$-JS divergences over training. This could be because even if the $\rm SNR^2$ distribution over the class having 5\% prevalence was not learnt well, it may not significantly impact the JS divergence \cite{teaching}.
% However, this is probably because the dataset is very lopsided and so skews heavily towards \textit{SND-2} later into training as seen in Fig. \ref{fig:SNR_2_us}. 
Finally, it can be seen that the the GAN struggles in this case to properly reproduce the direct-simulated $\rm SNR^2$ distributions. 
In the case of \textit{USS Mixed 50-50}, the $\rm SNR^2$ distributions of the two classes have greater variance for the GAN-generated images. This results in the GAN producing more images having a value of $\rm SNR^2$ intermediate between the two classes. For the \textit{USS Mixed 95-5} SIM, the GAN was not able to reproduce the mode corresponding to the class having 5\% prevalence in the dataset, as seen in \autoref{fig:SNR_2_us}. 

\begin{figure}
     \includegraphics[width=\linewidth]{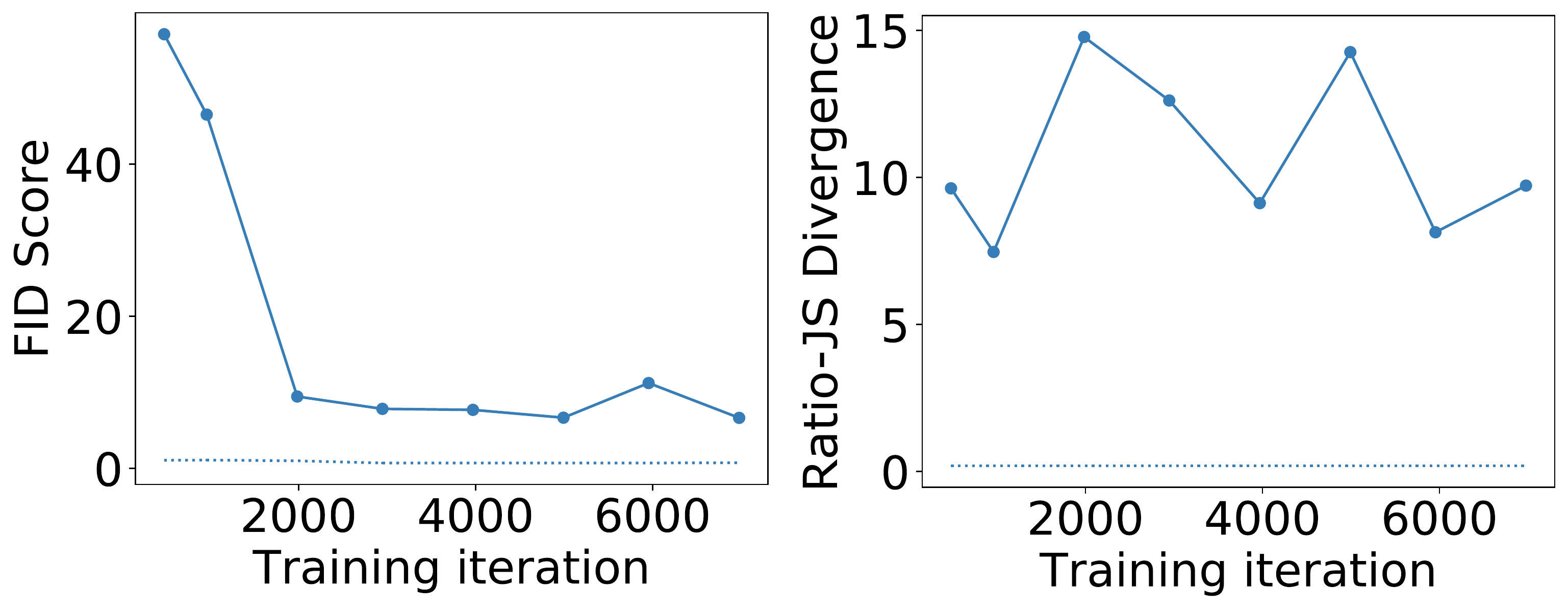}
      \caption{FID and empirical ratio-JS divergence between real and GAN-generated distributions for the S2V dataset. The dotted line represents the value of the measures between two direct-simulated datasets.}
      \label{fig:fid_js_victre}
\end{figure}

\begin{figure}
\includegraphics[width=\linewidth]{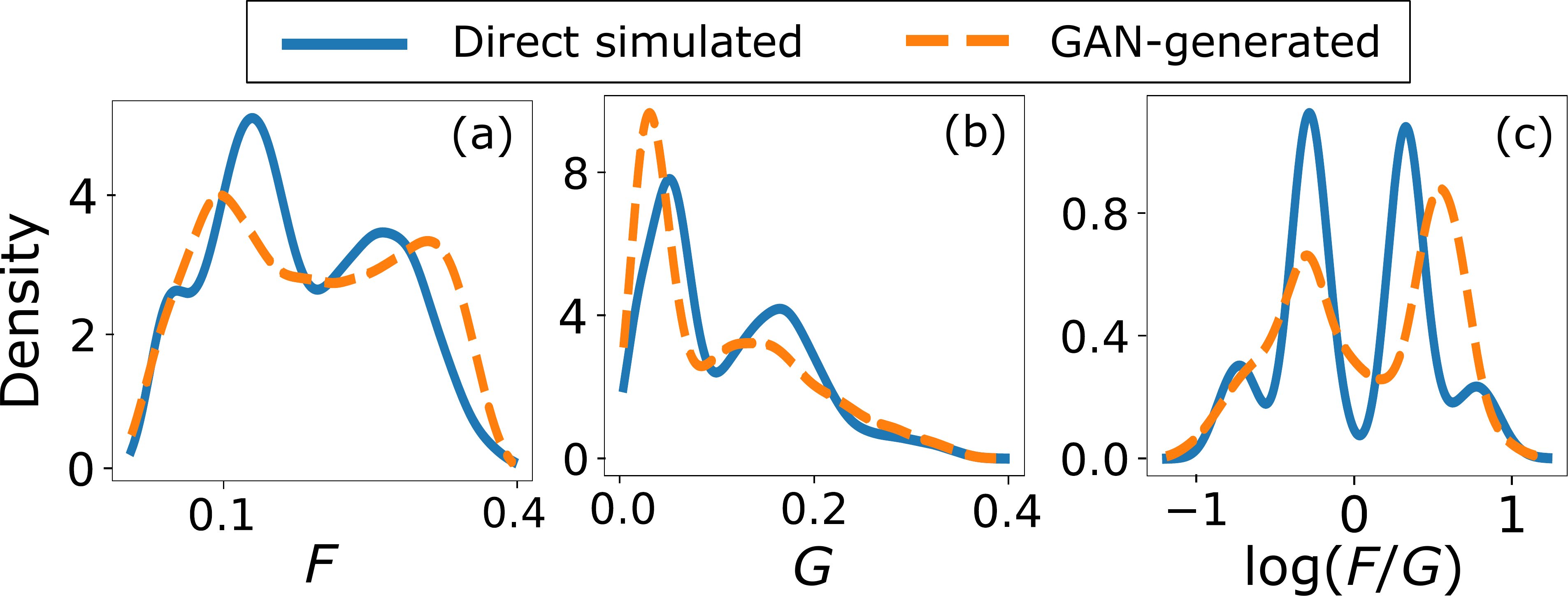}
\caption{(a-b) The estimated PDF over the per-image number of pixels corresponding to fat and glandular tissue respectively, as a fraction of the total image pixels (denoted by $F$ and $G$ respectively). (c) The estimated PDF over $\log(F/G)$.}
\vspace{-10pt}
\label{fig:fat_and_glandular_kde}
\end{figure}

\subsubsection{The S2V SIM}
Figure \ref{fig:fid_js_victre} shows the empirical JS divergence between the empirical PDFs of $\log\rho_{F:G}$ computed from the direct-simulated and GAN-generated images (henceforth refered to as the ratio-JS divergence) as a function of the training iteration. This is displayed alongside the plot of FID as a function of the training iteration. It can be seen that although the FID predictably converged as the training progresses, the ratio-JS divergence was erratic and did not converge the same way as FID. Figure \ref{fig:fat_and_glandular_kde} shows the empirical PDFs of $\log\rho_{F:G}$ computed on a per-image basis from the direct-simulated and GAN-generated images. The direct-simulated distribution clearly shows the four different breast types based on the $F:G$ ratio in their correct clinical prevalence. However, the GAN-generated distribution completely ignored or incorrectly represented many of the breast type modes. This was despite the GAN giving visually appealing images and accurate FID and other basic ensemble metrics.

\section{Summary}\label{sec:discussion}
Generative adversarial networks (GANs) could potentially be employed as stochastic image models for use in several tasks in medical imaging. However, GANs have traditionally been evaluated using mathematical or perceptual measures that may not correlate with those statistics that are important with respect to a downstream task. The objective of this work was to study the ability of GANs to reproduce medical image statistics that are meaningful and pertinent to the SIM under consideration, and to see how well traditional measures such as FID correlate with these pertinent statistics. 

The GANs employed consistently produced images that visually appeared realistic, and were able to accurately and consistently reproduce basic statistics such as the intensity histograms and image autocorrelation. It was also observed that although most of the evaluation measures used in this paper converged, they did not necessarily converge at the same rate, and some of them diverged as the training progressed. This indicates that the convergence of a commonly used measure such as the FID score to a low value does not guarantee the correct convergence of those statistics that are meaningful to the particular medical SIM under consideration. Since the FID score measures the Fr\'echet distance in the feature space of an Inception network trained on the ImageNet dataset, it is not tailored to the specific medical image distribution considered. Additionally, the GAN may learn the distribution of different features to different degrees of fidelity, resulting in different performance rankings when examined by different measures.

We note that for all the SIMs considered in this paper, the GAN-generated images retained potentially impactful per-realization errors in some of the meaningful features identified. These errors manifested themselves in the empirical distributions of these meaningful features learned by the GAN, where among others, critical inaccuracies such as mode-dropping and merging of multiple classes or modes was observed. This was despite the GAN producing excellent agreement with the direct-simulated distribution in terms of ensemble measures, such as the FID and basic first- and second-order statistics. 

These observations point to the need for choosing evaluation measures that are meaningful and pertinent to the SIM considered, are motivated by a downstream task, and are sensitive to the important aspects of a medical image distribution, such as multiple modes. While formulating such evaluation measures requires significant effort, it opens up the possibility of evaluating GANs in terms of those statistics that influence task-performance.

This study employed the StyleGAN2 architecture, since it has been shown to consistently produce realistic images when trained on a wide variety of datasets. However, the proposed analysis does not depend upon the GAN architecture employed, and could easily be performed on other GAN architectures. Canonical SIMs that produce simulated medical images provided the ability to examine the behavior of the GAN under a controlled setting with different parameter configurations. Nevertheless, evaluating GANs trained on real medical images remains a topic of future investigation. Lastly, although careful identification of meaningful evaluation measures is a key aspect of this study, it falls short of performing a task-based assessment of GANs. This will be the topic of a follow-up study.

% \section{Conclusion}\label{sec:conclusion}

%\section*{References}
%\label{sec:references}

\bibliography{refs}{}
\bibliographystyle{IEEEtran}

\end{document}

%\doublespacing

\title{Evaluating Procedures for Establishing Stochastic Image Models Using Generative Adversarial Networks -- Supplementary Information}

\author{A, B, C
\thanks{This work was supported in part by NIH Awards EB020604, EB023045, NS102213, EB028652, and NSF Award DMS1614305.}
\thanks{A is 
with the Department of Electrical and Computer Engineering, University of Illinois at Urbana-Champaign, Urbana, IL 61801 USA}
\thanks{B is with 
the Department of Electrical and Computer Engineering, University of Illinois at Urbana-Champaign, Urbana, IL 61801 USA}
\thanks{O is with 
the Department of Electrical and Computer Engineering, University of Illinois at Urbana-Champaign, Urbana, IL 61801 USA}
}
\maketitle

\begin{abstract}

\end{abstract}

\begin{IEEEkeywords}

\end{IEEEkeywords}

\section{Results - Ultrasound frequency sweep} \label{sec:freq_sweep}

\begin{figure}[h]
     \centering
     \includegraphics[width=0.97\linewidth]{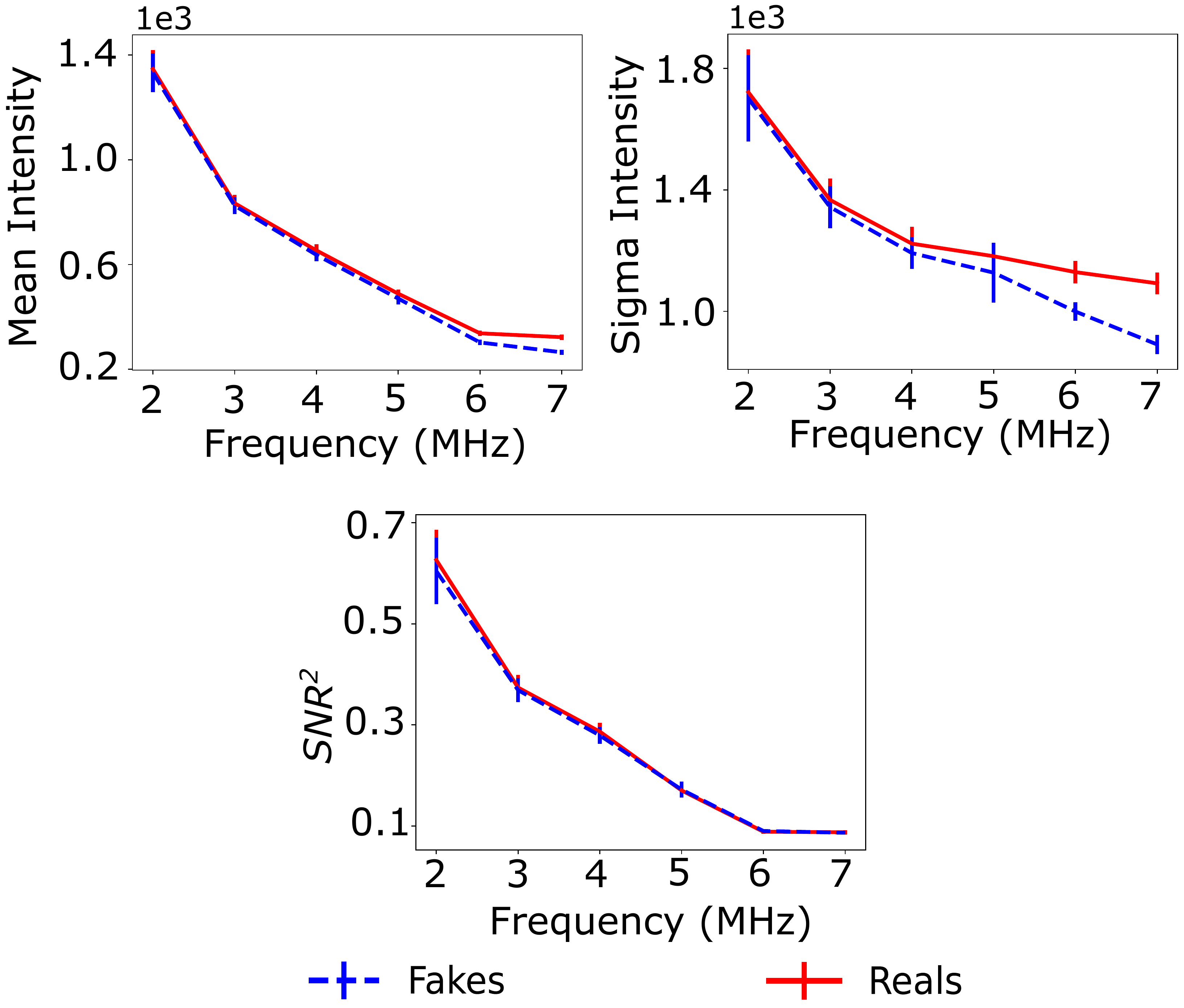}
      \caption{$SNR^2$, $\mu_I$ and $\sigma_I$ values for real and GAN generated images versus the carrier frequency.}
      \label{fig:freq_sweep_US}
\end{figure}

\begin{figure}
     \includegraphics[width=\linewidth]{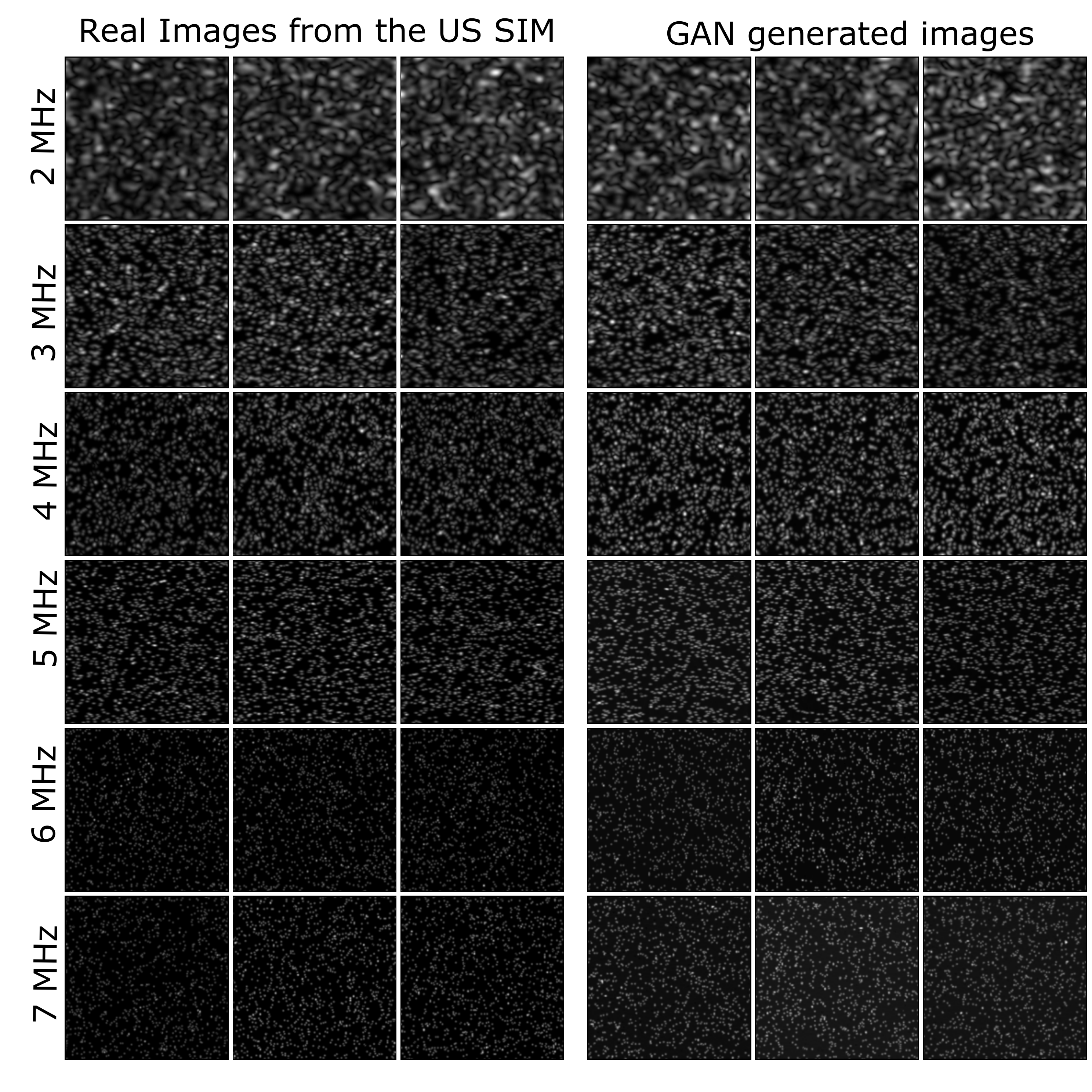}
      \caption{Real images from the US SIMs and fake images generated by the GANs for the SND-1-$f$ models.}
      \label{fig:real_and_fake_freq_sweep}
\end{figure}

\begin{figure*}
     \centering
     \includegraphics[width=\linewidth]{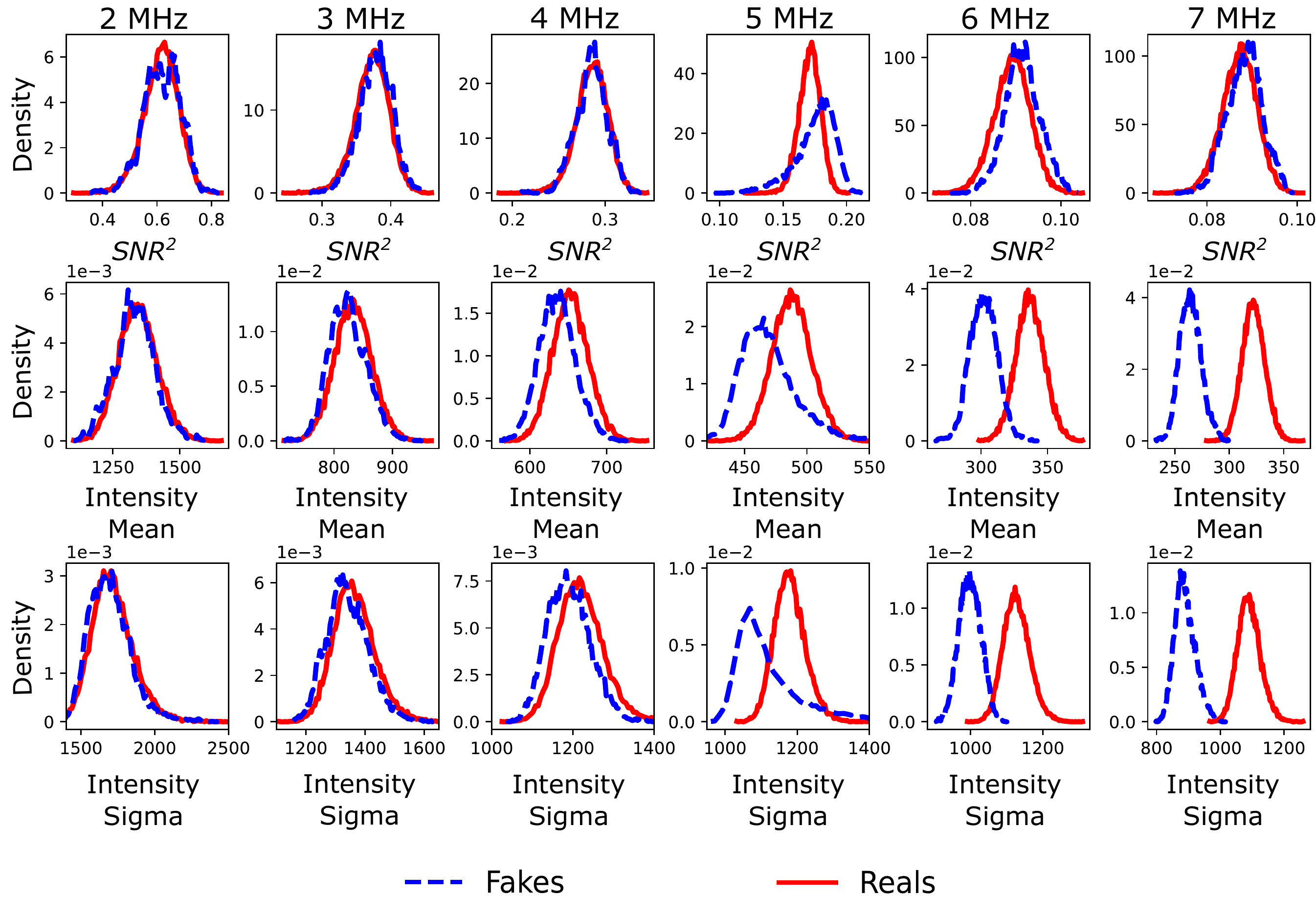}
      \caption{$SNR^2$, $\mu_I$ and $\sigma_I$ histograms for real and GAN generated ultrasound images with carrier frequencies 2,3,4,5,6 and 7 MHz.}
      \label{fig:SNR_2_freq_US}
\end{figure*}

Figure \ref{fig:real_and_fake_freq_sweep} shows GAN generated images as well as images from the training set each at different carrier frequencies. In Figure \ref{fig:freq_sweep_US} results for the intensity $SNR^2$, $\mu_I$ and $\sigma_I$ are plotted versus frequency for both GAN generated and training set images. The line indicates the mean of each of the statistics over the 20,000 images used for both `real' and `fake' images, while the bars indicate plus or minus one standard deviation for each statistics' histogram. In Fig. \ref{fig:SNR_2_freq_US} histograms of $SNR^2$, $\mu_I$ and $\sigma_I$ for all six carrier frequencies are shown.

Visually, both the GAN generated images and the training set images look very similar. However, this is not the case when evaluating the similarity with regards to the three statistics mentioned. Interestingly, the GAN generated images have a lower $SNR^2$ on average than the `real' images for frequencies less than $4$ MHz, but have larger $SNR^2$ values on average for frequencies greater than or equal to $4$MHz. A deeper investigation into the $\mu_I$ and $\sigma_I$ statistics shows that the GAN generated images severely underestimate both statistics with respect to the training set images. In general, the larger the frequency the more the `real' and `fake' distributions diverge to the point where at $6$ and $7$ MHz, these two statistic distributions no longer have any overlap between the `reals' and `fakes'.

%Insert Figures

%\section*{References}
%\label{sec:references}

\bibliography{refs}{}
\bibliographystyle{IEEEtran}